\documentclass[final,1p,times]{elsarticle}

\makeatletter
\def\ps@pprintTitle{%
  \let\@oddhead\@empty
  \let\@evenhead\@empty
  \let\@oddfoot\@empty
  \let\@evenfoot\@oddfoot
}
\makeatother

\usepackage{amsmath}
\usepackage{amssymb}
\usepackage{amsthm}
\usepackage{siunitx}
\usepackage{ulem}

\date{2025}
\begin{document}

\begin{frontmatter}
\title{Toward Efficient Electrokinetic Energy Conversion via Topographic Modulation of Electrical Conduction}

\author[inst1]{Austin Dick}
\author[inst1]{Kushal Iyyapareddy}

\author[inst1]{Aktaruzzaman Al Hossain}

\author[inst1,inst2,inst3]{Carlos E. Colosqui}
\ead{carlos.colosqui@stonybrook.edu}

\affiliation[inst1]{organization={Department of Mechanical Engineering, Stony Brook University},
            addressline={NY 11794}, 
            country={USA}}

\affiliation[inst2]{organization={Department of Applied Mathematics and Statistics, Stony Brook University},
            addressline={Stony Brook, NY 11794}, 
            country={USA}}

\affiliation[inst3]{organization={The Institute of Energy: Sustainability, Environment, and Equity, Stony Brook University},
            addressline={Stony Brook, NY 11794}, 
            country={USA}}

\begin{abstract}
%
This work presents experimental and theoretical analyses of electrokinetic flow in microchannels with glass and silica surfaces across a broad range of electrolyte concentrations (10$^{-5}$ to 10$^{-1}$ M). 
We demonstrate simple but effective strategies for controlling electrical conduction by engineering nanoscale and microscale topographic features that directly modify the structure and extent of the electric double layer (EDL) and the interfacial ion conduction pathway. 
These tailored surface topographies modulate the overall electrical conductivity in slit microchannels through similar phenomena documented for nanochannels and nanopores due to the presence of liquid-filled nanoscale topographic features with high concentration of highly mobile protons. 
The findings of this work reveal that the interaction between tailored surface features and the EDL can substantially enhance energy conversion efficiency in microscale systems. 
These insights along with simple analytical models provide guidance for the rational design and optimization of scalable electrokinetic devices and are broadly relevant to numerous energy harvesting and charge-separation technologies.
\end{abstract}



\end{frontmatter}



\section{Introduction}
\label{sec:intro}
Electrokinetic flow has been investigated as a mechanism for energy conversion and harvesting since the foundational studies by Osterle and co-workers in the 1960s~\cite{oldham1963streaming,osterle1964,morrison1965electrokinetic}. Research in this area has grown substantially over the past two decades, particularly in the context of micro- and nanofluidic systems~\cite{schoch2008,eijkel2005,pennathur2005}, and more recently through the development of engineered nanostructured surfaces and advanced nanomaterials~\cite{haldrup2016tailor,qin2020constant,aktar2020,wang2022mxenes}.
The growing interest in electrokinetic phenomena is largely motivated by their potential to support novel technologies for sustainable energy generation and storage, as well as for water purification and desalination.
Representative examples of such technologies include electrokinetic generators driven by capillary or evaporative flow~\cite{zhang2017capillary,jiao2023,liu2024energy}, triboelectric nanogenerators~\cite{yuan2024recent,zhao2024design,somton2025transparent}, salinity-gradient-based blue energy harvesting devices~\cite{iglesias2020combining,wang2024salinity}, and capacitive deionization systems~\cite{ma2019energy,younes2024review}, among others.
A key advantage of these technologies is their compatibility with abundant, low-cost materials and renewable resources. However, their broader viability, particularly in terms of scalability and commercial relevance, even for niche markets, relies on improving energy conversion efficiency.

Classical descriptions of electrokinetic flow in micro- or nanoscale channels consider a perfectly flat surfaces with a sharp (zero-thickness) liquid-solid interface that is homogeneously charged.
The classical description has been extensively adopted to rationalize experimental observations for a wide range of systems having nano- to macroscale characteristic dimensions.
On the other hand, electrokinetic phenomena in the presence of localized physical and chemical heterogeneity remain relatively understudied. 
Only a limited number of prior works have considered the effects of engineered nanoscale surface structures on the surface and zeta potentials \cite{grover2013preparation,haldrup2016tailor,aktar2020,huang2020molecular,goyal2024generalizing}.
The linear response theory for electrokinetics, based on Onsager's reciprocal relations~\cite{onsager1931reciprocal,miller1960thermo}, establishes that the steady-state energy conversion efficiency, whether the system operates as a generator or a pump~\cite{osterle1964}, is a monotonic function of the figure of merit $Z = \frac{(\epsilon \zeta / \mu)^2}{\kappa_H \kappa_E}$, which depends on the fluid permittivity $\epsilon$, viscosity $\mu$, zeta potential $\zeta$, hydrodynamic conductivity $\kappa_H$, and overall electrical conductivity $\kappa_E$ of the flow channel.
The maximum theoretical efficiency $\eta_{\text{max}} = \frac{(1 - \sqrt{1 - Z})^2}{Z}$ is reached when the external load resistance matches $R_L = \kappa_E^{-1} L / (A \sqrt{1 - Z})$, where $L$ and $A$ denote the channel length and cross-sectional area, respectively~\cite{xuan2006thermodynamic,van2006electrokinetic}.
As $\eta_{\text{max}}$ approaches unity when $Z \to 1$, improving electrokinetic efficiency involves increasing $\zeta$ and/or reducing $\kappa_H$ and $\kappa_E$. Surfaces with high surface charge, such as metal oxides and certain minerals, are favorable due to their hydrophilicity and minimal hydrodynamic slip, which lowers $\kappa_H$.

Alternatively, enhancing hydrodynamic slip through superhydrophobic surfaces~\cite{joly2004hydrodynamics,davidson2008electrokinetic,ren2008slip,papadopoulos2012electrokinetics} or drag-reducing polymers~\cite{berli2010electrokinetic,zhao2013electrokinetics,zhang2020soft} can also increase $\zeta$ and $\kappa_H$. However, these approaches may reduce liquid-solid interaction, diminishing surface charge and raising $\kappa_E$~\cite{joly2006liquid,heverhagen2016slip,aktar2020}.
Another approach to improve $\eta$ is to reduce $\kappa_E$ by lowering ion concentration, since electrical conductivity is mainly determined by mobile ions in solution. Reducing electrolyte concentration can lower $\kappa_E$ significantly, especially at $C \sim 10^{-4}$ to $10^{-2}$~M. In this regime, zeta potential tends to increase as surface charge decreases more slowly and the Debye length $\lambda_D$ increases proportionally to $C^{-1/2}$.
However, fundamental and practical limitations arise at vanishing $C$. To ensure finite entropy generation, $Z < 1$ must hold, which imposes a minimum conductivity
$\kappa_E^0 = \frac{(\epsilon \zeta / \mu)^2}{\kappa_H}$.
Given $\kappa_H \propto H^2$ for a slit channel of height $H$, this leads to $\kappa_E \ge \kappa_E^0 \propto H^{-2}$ for finite $\zeta$ as $C \to 0$.
Thus, an irreducible conductivity must arise from mechanisms not captured in classical models assuming sharp interfaces, flat surfaces, immobile Stern layer ions, and constant material properties.

Numerous experiments have reported unexpectedly high and nearly constant conductivities ($\kappa_E \sim 10^{-3}$ to $10^{-1}$~S/m) in silica nanochannels at low electrolyte concentrations ($< 10^{-4}$~M)~\cite{stein2004surface,schoch2005ion,ritt2022thermodynamics,lee2012large,lin2021surface,zhan2023ion,aluru2023fluids}.
This study shows that similar excess surface conductivity also emerges in microscale channels at low ionic strength, a phenomenon previously attributed only to nanoscale confinement.
We propose that mobile charges accumulate in nanoscale topographic features at the interface, leading to enhanced conductivity.
We also demonstrate that modifying the surface structure and channel geometry allows for control over this interfacial conductivity.
Finally, we introduce an analytical model that agrees with experiments and provides a practical framework for designing electrokinetic systems with improved energy conversion efficiency.

\section{Theoretical Description \label{sec:theory}}
\begin{figure}[h!]
\centering
    \includegraphics[width=0.7\columnwidth]{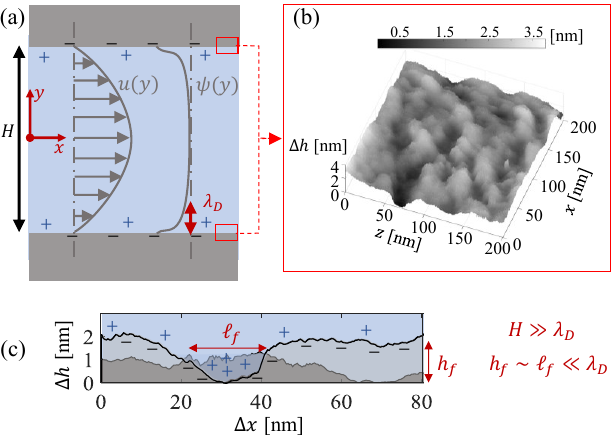}
    \caption{Electrokinetic flow in a slit microchannel with height $H \gg \lambda_D$.  
(a) Flow and electric potential distributions based on the classical sharp-interface model, assuming perfectly flat, hydrophilic, and negatively charged surfaces.  
(b) Three-dimensional nanoscale surface topography of a real substrate, illustrated by the topographic height map of a borosilicate glass surface obtained via atomic force microscopy.  
(c) Superimposed one-dimensional topographic height profiles revealing nanoscale features smaller than the Debye length. These liquid-filled nanocavities and grooves, characterized by $h_f \sim \ell_f \ll \lambda_D$, promote local accumulation of protons and counterions within the electric double layer.} 
\label{fig:channel}
\end{figure}

Analytical developments presented in this section are tailored to the case of 1:1 symmetric electrolytes confined within a slit channel featuring symmetric walls and a height $H \gg \lambda_D$, where the channel dimensions are significantly larger than the Debye screening length $\lambda_D$ (Fig.~\ref{fig:channel}a).
Within the classical theoretical framework, which treats the liquid-solid boundary as ideally sharp and planar, we extend the model to account for the influence of three-dimensional nanoscale surface roughness. This includes random height variations $\Delta h$ (Fig.~\ref{fig:channel}b), as well as more structured topographic features defined by a characteristic height $h_f$ and lateral dimension $\ell_f$.
These nanoscale features, which form on the charged channel surfaces, are much smaller than the overall channel height ($h_f \sim \ell_f \ll H$), and at moderate to low electrolyte concentrations ($C \lesssim 1$~mM), are also much smaller than the Debye length $\lambda_D$ (Fig.~\ref{fig:channel}c).

The classical description of steady electrokinetic flow yields the volumetric flow rate $Q$ and electric current $I$ produced by a pressure difference $\Delta p$ and voltage difference $\Delta V$ across a channel of constant cross-sectional area $A$ and length $L$ (cf. Fig.~\ref{fig:channel}a), as the linear system
\begin{equation}
\begin{bmatrix}
Q/A\\
I/A
\end{bmatrix}
=
\begin{bmatrix}
\kappa_H & - \frac{\epsilon \zeta}{\mu}\\
- \frac{\epsilon \zeta}{ \mu} & \kappa_E
\end{bmatrix}
\begin{bmatrix}
{\Delta P/L}\\
{\Delta V/L}
\end{bmatrix},
\label{eq:matrix}
\end{equation}
where the coupling between fluid flow and charge transport is determined by the electrokinetic zeta potential $\zeta$, which is a function of the surface potential $\psi_o=\psi_o(\sigma)$ for a given surface charge density $\sigma$, electrolyte concentration, and solution pH.
We will consider charge-regulating surfaces that acquire net charge through protonation/deprotonation reactions. We thus adopt a conventional 2-pK model to determine the concentration-dependent surface charge density \cite{healy1978} 
%
\begin{equation}
\sigma(\psi_o)=e \Gamma 
\frac{ 2\sinh[(\psi_N-\psi_o)e/k_B T] 10^{(pH_0-pK_{-})} }
{1+2\cosh[(\psi_N-\psi_o)e/k_B T] 10^{(pH_0-pK_{-})}},
\label{eq:charge}
\end{equation}
where $\Gamma$ is the surface density of chargeable sites, $pK_{\mp}$ are the proton dissociation/adsorption constants, $pH_0=(pK_+ + pK_-)/2$ 
is the isoelectric point pH, and $\psi_N=-2.3 (k_BT/e) (\mathrm{pH}-pH_0)$ is the so-called Nernst potential.

Adopting the classical Stern-Gouy-Chapman description, the EDL is conformed by a Stern layer of adsorbed water molecules and immobilized counterions and a diffuse layer of mobile ions that can be advected by the local flow.
The diffuse-layer potential at the outer boundary of the Stern layer is  
\begin{equation}
\psi_d=\psi_o-\sigma/C_s,  
\label{eq:stern}
\end{equation}
where $C_s$ is the specific Stern capacitance.
For an electroneutral system one must satisfy the Grahame equation \cite{grahame1947electrical}, which for the case of a 1:1 symmetric electrolyte and $H\gg\lambda_D$ is 
\begin{equation}
\psi_d=2 \frac{k_B T}{e} \mathrm{arcsinh}\left(\frac{\sigma e \lambda_D}{2\epsilon k_B T}\right).
\label{eq:grahame}
\end{equation}
The Debye length, $\lambda_D=\sqrt{\epsilon k_BT/2 e^2 C N_A}$, is determined by the electrolyte molar concentration $C$, where $k_B$ is the Boltzmann constant, $T$ is the system temperature, $e$ is the elementary charge, and $N_A$ is the Avogadro number. 
The surface charge $\sigma$ and potential $\psi_o$ are determined by solving numerically the set of Eqs.~\ref{eq:charge}--\ref{eq:grahame}.

\subsection{Electrokinetic potential and electrical conductivity \label{sec:electrokinetics}} 
To derive analytical expressions for the zeta potential, hydrodynamic conductivity, and electrical conductivity, we adopt the classical framework for a perfectly flat channel, which neglects nanoscale topographic height variations $\Delta h \ll H$.
Within this framework, the electrostatic potential is obtained from the Poisson-Boltzmann (P-B) equation, $\nabla^2 \psi = -\rho_E/\epsilon$, with the charge density given by $\rho_E = -2ecN_A \sinh(e\psi/k_B T)$. This yields the solution
\begin{equation}
\psi(y) = 4\mathrm{arctanh}\left[\tanh\left(\frac{e\psi_d}{4k_B T}\right) e^{\frac{|y|-H/2}{\lambda_D}}\right],
\label{eq:psi}
\vspace{-6pt}
\end{equation}
for a slit channel with symmetric walls.
The electrokinetic velocity profile is then obtained by solving the steady-state Navier-Stokes equation, $\mu \nabla^2 u = \epsilon \nabla^2 \psi \cdot \Delta V/L - \Delta p/L$, which gives:
\begin{equation}
u(y) = \frac{\epsilon \Delta V}{\mu L} \left[\psi(y) - \psi_d\right] + \frac{\Delta P}{2\mu L} \left[\left(\frac{H}{2}\right)^2 - y^2\right].
\label{eq:flow}
\end{equation}

The zeta potential $\zeta$ for hydrophilic slit channels can be inferred from the electroosmotic flow profile $u_E = u(y)$ with $\Delta p = 0$ as:
\begin{equation}
\zeta = -\frac{\mu L}{\epsilon \Delta V} \cdot \frac{2}{H} \int_0^{H/2} u_E(y) \, dy = \psi_d - \bar{\psi},
\label{eq:zeta_th}
\end{equation}
where the average electrostatic potential is defined as $\bar{\psi} = (2/H) \int_0^{H/2} \psi(y) \, dy \sim \mathcal{O}(\lambda_D/H)$.
Similarly, the hydrodynamic conductivity is determined from the Poiseuille flow profile $u_P = u(y)$ with $\Delta V = 0$, leading to the standard result:
\begin{equation}
\kappa_H = \frac{2}{H} \int_0^{H/2} u_P(y) \, dy = \frac{H^2}{12\mu},
\label{eq:kH}
\end{equation}
which holds under steady-state flow in slit channels.

We consider three parallel ion conduction pathways: through the bulk electrolyte, the electric double layer (EDL), and the liquid-filled nanoscale interfacial topography (cf. Fig.~\ref{fig:gaps}a). Thus, the total electrical conductivity is expressed as:
\begin{equation}
\kappa_E = \kappa_B + \kappa_{EDL} + \kappa_{ex},
\label{eq:cond_th}
\end{equation}
where the surface contribution, or EDL conductivity, is given by:
\begin{equation}
\kappa_{EDL} = \frac{4e\epsilon C N_A}{\mu H} \int_0^{H/2} \sinh\left(\frac{e\psi}{k_B T}\right) [\psi_d - \psi(y)] \, dy,
\label{eq:kEDL}
\end{equation}
and is computed using Eqs.~\ref{eq:psi}--\ref{eq:flow} by evaluating ion transport driven by electroosmotic flow $u_E$, assuming a flat, sharp-interface geometry.
In this regime, both bulk and surface conductivities scale with ion concentration, and the EDL conductivity $\kappa_{EDL} \ll \kappa_B$ remains negligible in microscale channels where $H \gg \lambda_D$.

The third term in Eq.~\ref{eq:cond_th}, $\kappa_{ex}$, represents an ``excess'' conductivity that accounts for charge transport mechanisms not captured by bulk or EDL contributions. These include ion movement through liquid-filled nanoscale surface features, such as cavities or grooves, where enhanced local electric fields and elevated proton mobility lead to increased conduction, as observed in systems with strong nanoscale confinement~\cite{duan2010anomalous,paiva2019conduction,ji2021electric,allemand2023anomalous,zhan2023ion}.
This excess term $\kappa_{ex}$ ensures finite entropy production as $C \to 0$ (see Sec.~\ref{sec:intro}) and is expected to dominate the overall conductivity at very low concentrations.
In this work, $\kappa_{ex}$ is treated as an effective material property, which can be extracted from experimental data at low ionic strengths and predicted analytically based on different interfacial topographies.

\subsection{Topographic control of electrical conduction \label{sec:topography}}
\begin{figure}[h!]
\centering
    \includegraphics[width=0.7\columnwidth]{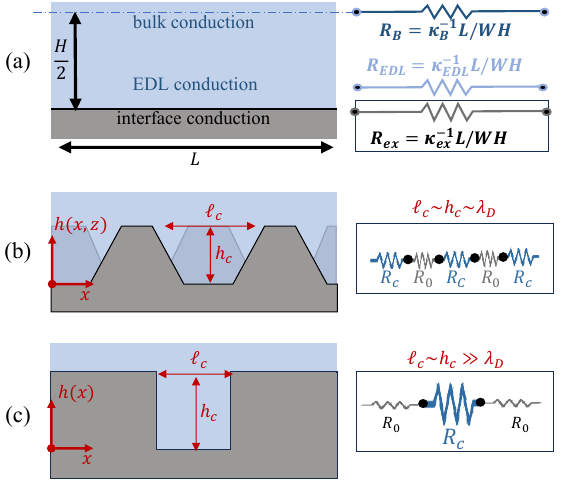}
\caption{Topographic control of electrical conduction at the solid-liquid interface. 
(a) Schematic of conduction pathways and associated electrical resistances: 
(i)bulk conduction with $\kappa_B=R_B^{-1}L/WH$, 
(ii) EDL or ``surface'' conduction with $\kappa_{EDL}=R_{EDL}^{-1}L/WH$, and 
(iii) excess interfacial conduction with $\kappa_{ex}=R_{ex}^{-1}L/WH$.
(b) Nanoscale cavities and interfacial resistance for characteristic dimensions comparable to the EDL thickness $\ell_C\sim h_C~\sim\lambda_D$.
(c) Microscale cavity and interfacial resistance for characteristic dimensions much larger than the EDL thickness $\ell_C\sim h_C~\gg\lambda_D$. 
\label{fig:gaps}}
\end{figure}

The theoretical framework presented in Sec.~\ref{sec:electrokinetics} for a slit channel with planar hydrophilic surfaces accounts for three distinct electrical conduction pathways (cf. Fig.~\ref{fig:gaps}a): conduction through the bulk electrolyte with conductivity $\kappa_B$, surface conduction within the electric double layer (EDL) with conductivity $\kappa_{EDL}$, and interfacial conduction through liquid-filled nanoscale surface features, characterized by an excess conductivity $\kappa_{ex}$.
At very low electrolyte concentrations, where $\kappa_B \to 0$ and $\kappa_{EDL} \to 0$, the interfacial conduction path, associated with an effective resistance $R_{ex} = \kappa_{ex}^{-1} L / WH$, is expected to dominate the total electrical conduction and therefore impose a fundamental limit on electrokinetic energy conversion efficiency.

To mitigate this interfacial contribution, we explore the use of ``large'' liquid-filled cavities incorporated into the channel walls. These cavities have characteristic height and lateral dimensions comparable to or exceeding the Debye length, as shown in Fig.~\ref{fig:gaps}b-c.
The proposed strategy is analogous to placing series resistive elements with an electrical resistance $R_C$ in the interfacial conduction pathway (cf. Fig.~\ref{fig:gaps}b-c), which results in a modulated electrical conductivity
\begin{equation}
\kappa_E=\kappa_{B}+\kappa_{EDL}
+\left[\kappa_{C}^{-1} \phi_C+\kappa_{ex}^{-1} (1-\phi_C)\right]^{-1} \frac{2h_C}{H},
\label{eq:kcontrol}
\end{equation}
where $\kappa_{C}$ is the conductivity of the liquid-filled cavity and $\phi_C$ is the fraction of channel surface area covered by such cavities.

\paragraph{Cavity conductivity}
In this study, we evaluate the proposed approach for reducing electrical conductivity by implementing two types of surface modifications:  
(i) channels with engineered nanostructured surfaces that incorporate uniformly distributed liquid-filled nanocavities with dimensions on the order of the Debye length (Fig.~\ref{fig:gaps}b), and  
(ii) channels featuring a single, isolated cavity with microscale dimensions significantly larger than the Debye length (Fig.~\ref{fig:gaps}c).
The conductivity of the liquid-filled cavity is estimated using the expression
\begin{equation}
\kappa_C = \mu_+ |\sigma| e^{-\frac{\psi_D}{k_B T}} \times \frac{S_C}{V_C},
\label{eq:kC}
\end{equation}
where $\mu_+ = 3.63 \times 10^{-7}$~m$^2$/V~s is the proton mobility, and the surface-to-volume ratio is approximated as $S_C / V_C \simeq \sqrt{1 + 4h_C^2 / \ell_C^2}$ for nanocavities (see Fig.~\ref{fig:gaps}b), and  
$S_C / V_C = 1 + 2h_C / \ell_C$ for the rectangular microscale cavity (see Fig.~\ref{fig:gaps}c).
The analytical model in Eq.~\ref{eq:kC} captures the contribution of highly mobile protons within the electric double layer of the cavity and offers an accurate approximation in the limit of low electrolyte concentration.
Comparisons between predictions from Eqs.~\ref{eq:kcontrol}--\ref{eq:kC} for both surface modifications (Fig.~\ref{fig:samples}c-d) and experimental results are presented in Sec.~\ref{sec:results}.

\section{Experimental Methods \label{sec:Exp}}

\subsection{Surface substrates and topography}
\begin{figure}[h!]
   \centering
    \includegraphics[width=1\textwidth]{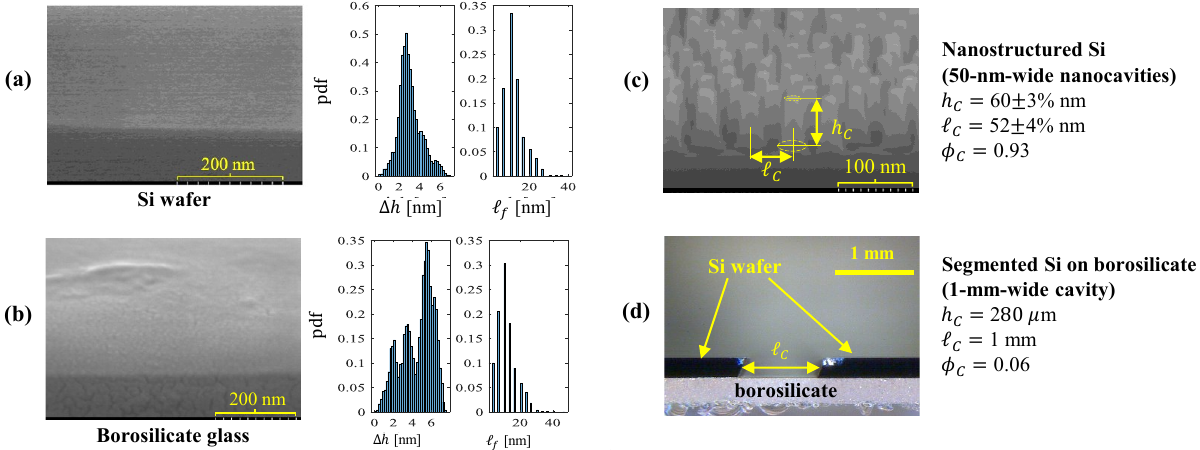}
\caption{Surface samples employed for the four studied flow channels. 
SEM images and characteristic dimensions of topographic surface features for:
(a) Mirror-polished Si wafers with native silica film,
(b) Uncoated borosilicate glass slides,
(c) Nanostructured Si surfaces, and
(d) Segmented Si wafers on borosilicate glass separated by a single rectangular cavity.
Panels (a) and (b) include the probability distribution function (pdf) for the random topographic height $h_f$ and peak-to-peak distance $\ell_f$.
Panels (c) and (d) report the average and nominal cavity height and lateral dimensions.}
\label{fig:samples}
\end{figure}

The surface substrates used in this work are silicon wafers (University Wafers, type N, phosphorus-doped, $<$100$>$ orientation, 275 $\mu$m thickness) and uncoated borosilicate glass slides (McMaster-Carr), which are composed of 70-87\% SiO$_2$ (see Fig.~\ref{fig:samples}a-b). 
The silicon substrates develop native SiO$_2$ films with nanoscale thickness upon exposure to ambient air or aqueous environments and exhibit similar Young contact angles ranging from 40$^\circ$ to 70$^\circ$ \cite{dhiraj2020,zhen2021,aktar2020,aktar2022}.
For borosilicate glass surfaces, the contact with water promotes the formation of silanol groups that enhance surface change and hydrophilicity.
The hydrophilic properties of the SiO$_2$ and borosilicate result in the formation of well wetted surfaces with liquid-filled topographic features of nano- and microscale dimensions \cite{dhiraj2020,zhen2021,aktar2020,aktar2022}. 

Topographic analysis performed with a profilometer (Dektak 150 Stylus) indicates that the mirror-polished wafers exhibit near-ideal flatness, with mean height deviations from flatness below 5 nm over distances of 1 cm. In contrast, the borosilicate glass surfaces employed display larger features with microscale dimensions between 1 to 5 $\mu$m.
Both substrates, however, exhibit similar nanoscale surface roughness with topographic height fluctuations with rms fluctuation $h_f\simeq 4$ nm and nanoscale lateral dimensions $h_f\simeq$~1-20 nm (see Fig.~\ref{fig:samples}a-b). 
Electrical conductivity measurements of the plain Si wafers report an solid-phase intrinsic conductivity of $16.1\pm 10\%$ S/m for a wide range of current (1-100 $\mu$A) and voltages (0.01-50 mV). 
These values fall within the range of conductivities reported by the supplier for the p-doped wafers used.
The conductivity of borosilicate glass surface is extremely low ($\lesssim 10^{-12}$~S/m) and falls below the minimum measurable conductivity of the employed probe ($10^{-4}$~S/m).

In addition, this study employs two topographic modifications of the channel surfaces with silicon substrates: 
(i) nanostructured silica (Fig.~\ref{fig:samples}c) with a regular array of conical pillars and nanocavities of height $h_C\simeq$~60 nm and period $\ell_C\simeq$~50 nm, the cavities occupy a surface fraction $\phi_C=0.93$,
(ii)segmented plain silicon on borosilicate (Fig.~\ref{fig:samples}d) separated by a single 1-mm-wide cavity, the cavity occupies a surface fraction $\phi_C=0.06$. 
The nanostructured surfaces (Fig.~\ref{fig:samples}c) fabrication process via block copolymer self-assembly and the topographic parameters are described in detail in previous work by our group \cite{aktar2020,aktar2022}.

\subsection{Electrokinetic flow apparatus}
\begin{figure}[h!]
    \centering
    \includegraphics[width=0.7\columnwidth]{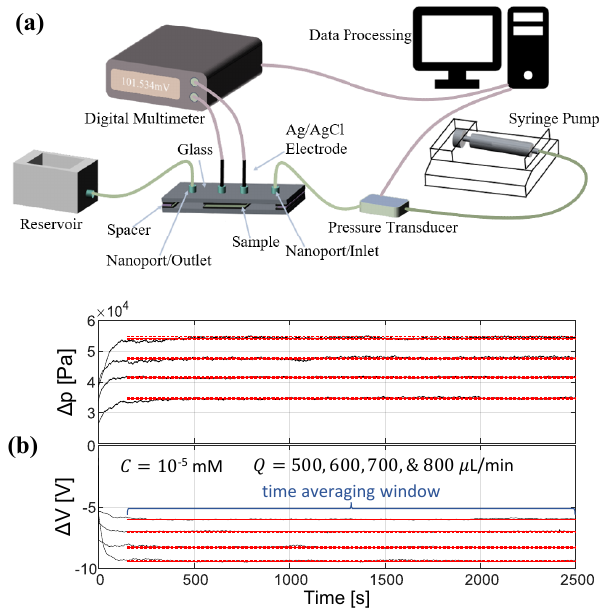}
    \caption{Experimental apparatus for electrokinetic flow characterization.
(a) Schematic of the flow cell and experimental apparatus.
(b) Representative pressure-time and voltage-time traces simultaneously acquired to determined average pressure and voltage difference under steady conditions at a prescribed flow rate for each electrolyte concentration. The conditions reported in panel (b) correspond to the borosilicate glass channel and DI water at four different flow rates $Q=$~500-800~$\mu$L/min.} 
    \label{fig:setup}
\end{figure}

Our experimental study utilizes a custom-built flow cell and measurement apparatus (Fig.~\ref{fig:setup}a) consisting of two large reservoirs connected by a slit microchannel with suitable dimensions (height $H=25$~$\mu$m, length $L=3.81$~cm, and width $W=2.54$~cm) for precise measurement of voltage differences, hydrodynamic conductance, and electrical conductances across a wide range of electrolyte concentrations.
The flow cell design enables the study of different surface samples and channel configurations to control electrical conductivity as discussed in Sec.~\ref{sec:topography}.
Four specific configurations are examined in this study:
(A) Channel with plain Si surfaces (Fig.~\ref{fig:samples}a),
(B) Channel with nanostructured Si surfaces (Fig.~\ref{fig:samples}b),
(C) Channel with segmented plain Si surfaces on borosilicate (Fig.~\ref{fig:samples}c), and
(D) Channel with plain borosilicate surfaces (Fig.~\ref{fig:samples}d).

The flow cell enclosure consists of two borosilicate glass slides separated by spacers (Fig.\ref{fig:setup}a). 
The surface samples forming the slit microchannel are attached to the glass slides using a low-viscosity epoxy, the flow cell is fully sealed with waterproof epoxy.   
As illustrated in Fig.~\ref{fig:setup}a, the flow cell has inlet/outlet ports separated at a distance $L_p=5.08$~cm and two additional ports separated by a distance $L_v=2.54$~cm through which AgCl pellets (1.0 mm in diameter and 2.0 mm in length) are inserted.
In all the studied configurations the electrodes are positioned inside the flow channel to eliminate undesirable experimental uncertainties that can arise when measuring voltage differences in the inlet/outlet reservoirs, e.g., due to concentration polarization and deviations from unidirectional fully developed flow at the channel-reservoir junction.

Four PEEK NanoPorts (from IDEX with size 10-32 coned threading) are bonded to the top surface of the flow cell. Tubing is attached to the inlet and outlet ports and the inlet port is connected to the syringe pump (Chemyx Fusion 200) supplying the electrolyte solution. 
A multimeter (Keithley DMM-6500) with high impedances up to 10 G$\Omega$ is employed to obtain voltage and resistance measurements.
An in line pressure sensor (IDEX I2C PS200) placed between the inlet port and the syringe pump (Fig.~\ref{fig:setup}a) is employed to take differential pressure measurements.
The pressure and voltage measurements are simultaneously recorded as the electrolyte solution is forced to flow at a prescribed volumetric rate.

\subsection{Measurement procedure}
Sodium chloride (NaCl, Sigma-Aldrich, 99.5\%) solutions in deionized (DI) water (Lab Chem, ASTM Type I) at five different concentrations 
$C = 0, 0.1, 1, 10, 50$ mM are employed for the experimental analysis. 
The measured solution pH ranged between 6 and 6.5 for all studied cases.
For the studied conditions, the measured bulk conductivities closely match those predicted by the analytical expression
$\kappa_B=\Lambda_M C+ N_A\mu_{+} 10^{-pH+3}$, where $\Lambda_M= 12.64 \, \text{S} \cdot \text{m}^2/\text{mol}$ is the molar conductivity for fully dissolved Na$^+$ and Cl$^-$ ions, and $\mu_+=3.63\times 10^{-7}$~m$^2$/V~s is the proton mobility in water due to the Grotthuss mechanism \cite{agmon1995grotthuss,kornyshev2003,popov2023}. 
The bulk conductivity measured for DI water (0.11~mS/cm) corresponds to a molar concentration of 0.0088 mM NaCl, and therefore this value is used to report results for electrokinetic flow measurements conducted with DI water.
To enhance reproducibility, the AgCl electrodes are removed and cleaned prior to experimental measurements by immersing them in ethyl alcohol for 60 seconds, followed by rinsing with DI water. 
To remove any trapped air, the ports are exposed to the atmosphere while flushing the system with DI water. 

For each solution concentration, digital data for pressure and voltage are collected during 2500 s at four different flow rates $Q=$~300-800~$\mu$L/min, with at least two replicate runs conducted per concentration.
As shown in Fig.~\ref{fig:setup}b, the system requires 100--200 s to reach steady-state conditions. After this period, the time-averaged pressure head $\Delta p$ and voltage difference  $\Delta V$ are recorded for each flow rate. Relative uncertainties are below 5\% for all reported cases.
Under steady-state conditions, the time-averaged pressure head $\Delta p$ increases linearly with flow rate, consistent with analytical predictions for Poiseuille flow, and confirming a nominal channel height of $H = 25 \pm 8\%$~ $\mu$m.
The measured average voltage difference $\Delta V$ is linearly proportional to $\Delta p$ and the electrokinetic coupling coefficient (streaming voltage-to-pressure slope)  
$dV/dp = (\Delta V/\Delta p) \times (L_p/L_v)$ thus is determined from a linear regression of the experimental data.

\section{Results and discussion \label{sec:results}}
This section presents electrokinetic flow measurements obtained using the materials and experimental procedures described in Sec.~\ref{sec:Exp}, along with analytical predictions developed in Sec.~\ref{sec:theory} using the model parameters listed in Table~1. 
Four channel configurations are studied: (Channel A) plain Si, (Channel B) nanostructured Si with 50-nm-wide nanocavities, (Channel C) segmented plain Si with a single 1-mm-wide cavity, and (Channel D) bare borosilicate glass (cf. Fig.~\ref{fig:channel}a-d).
By comparing experimental data with theoretical predictions, we quantify how the different surface topography modifications influence electrical conduction, particularly at low electrolyte concentrations.

\begin{table}[hbt!]
\centering
\begin{tabular}{ p{3. cm}|c|c|c}
 \hline
Channel (substrate)  & $\Gamma$ [nm$^{-2}$] &  $pK_-$ & $\kappa_{ex}$ [S/m] \\
 \hline
A (plain Si)	& 5.0  &      7.8 &   9$\times 10^{-3}$ \\
B	(nanostructured Si) & 5.0  &      7.8 &   9$\times 10^{-3}$ \\
C	(segmented Si) & 5.0    &     7.8 &  9$\times 10^{-3}$ \\
D	(borosilicate) & 7.0    &      8.2&   7$\times 10^{-5}$ \\
\hline
\end{tabular}
\vskip 5pt
\caption{Parameters employed in the analytical predictions for the flow channels composed of Si and borosilicate glass substrates, corresponding to the four configurations illustrated in Fig.~\ref{fig:samples}. 
For all cases the isoelectric point is $pH_0=2.4$ and the Stern capacitance is $C_e$ [mF/m$^2$].}
\label{tb:phil}
\end{table}

\begin{figure*}[hbt!]
    \centering
    \includegraphics[width=1\textwidth]{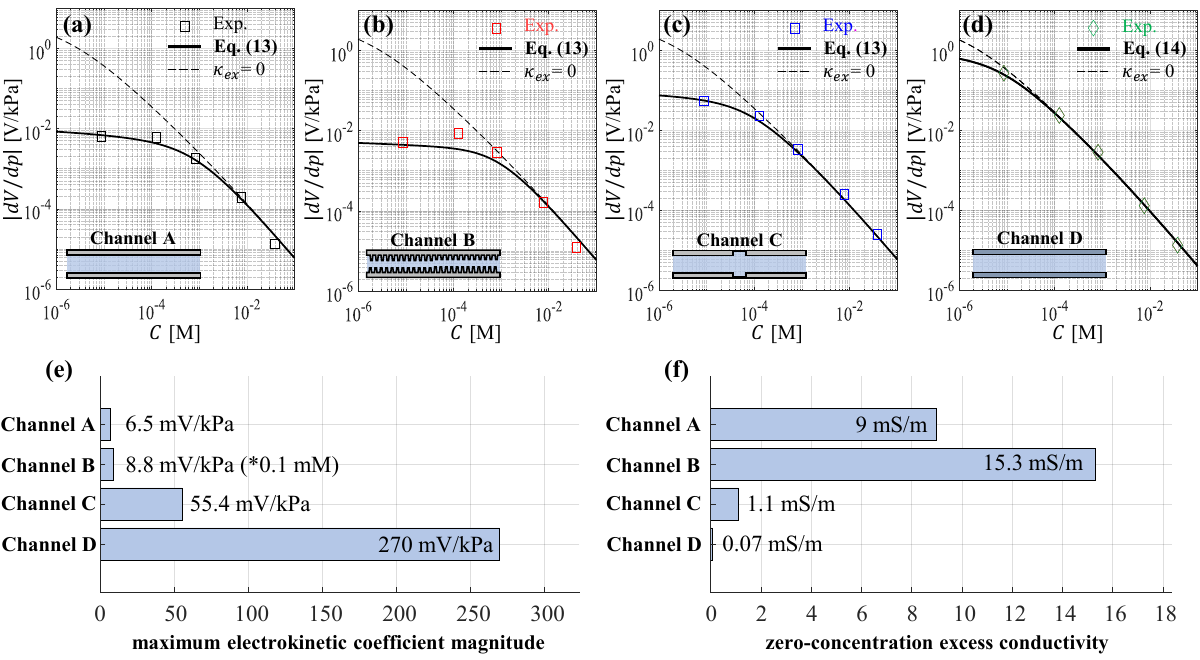}
\caption{Experimental and analytical results for the four studied channels: 
(Channel A) Plain Si, 
(Channel B) Nanostructured Si (50-nm-wide cavities), 
(Channel C) Segmented Si (single 1-mm-wide cavity), and 
(Channel D) Bare borosilicate slides. 
(a-d): Electrokinetic coefficient magnitude $|\Delta V/\Delta p|$ vs. NaCl concentration $C$.
Markers represent experimental measurements, black solid lines indicate analytical predictions from Eq.~\ref{eq:dpdV} with model parameters reported in Table 1, and dashed lines are predictions from Eq.~\ref{eq:dpdV} assuming zero excess conductivity ($\kappa_{ex} = 0$).
(e) Maximum electrokinetic coefficient magnitude attained for each studied channel: the maximum pressure-to-voltage conversion is observed for DI water except for Channel B (see legend and main text).  
(d) Zero-concentration excess interfacial conductivity determined by Eq.~\ref{eq:cond_th} (Channel A \& D) and Eq.~\ref{eq:kcontrol} (Channel B \& C).}
\label{fig:dVdP}
\end{figure*}

\paragraph{Zeta potential and surface charge} 
The zeta potential (Eq.~\ref{eq:zeta_th}) is computed from the concentration-dependent surface charge determined by the 2-pK model (Eqs.~\ref{eq:charge}--\ref{eq:grahame}), using site densities of $\Gamma =$~5~nm$^{-2}$ for Si and $\Gamma =$~7~nm$^{-2}$ for borosilicate substrates. A common isoelectric point of $pH_0=2.4$ and Stern capacitance $C=0.2$ F/m$^2$ are applied to all studied cases (cf. Table~1).
The dissociation constant $pK_{-}=7.8$ is adopted for the Si substrates and $pK_{-}=8.2$ is employed for the borosilicate glass.
This parameter set, consistent with prior studies predicts negative surface charges in the range $\sigma \simeq$~ -10 to -1~mC/m$^2$ for near neutral pH values, with weaker charge at lower concentrations \cite{behrens2001charge,barisik2014size,yang2020understanding,aktar2020}. 
%

\subsection{Electrokinetic coupling coefficient}
The analytical expression for the electrokinetic coupling coefficient is given by
\begin{equation}
\frac{dV}{dp} = \frac{\epsilon_0 \epsilon_R}{\mu} \cdot \frac{\zeta}{\kappa_E},
\label{eq:dpdV}
\end{equation}
where the zeta potential $\zeta$ is obtained from Eq.~\ref{eq:zeta_th}, and the electrical conductivity $\kappa_E$ is determined using Eq.~\ref{eq:cond_th} for channels with planar surfaces (Channels A and D), and Eq.~\ref{eq:kcontrol} for those with modified surface topography (Channels B and C).

In all cases, the electrokinetic coupling coefficient $dV/dp$ is negative, and its magnitude is reported in Fig.~\ref{fig:dVdP}, along with theoretical predictions based on Eq.~\ref{eq:dpdV}. These predictions include both the limiting case of zero excess conductivity ($\kappa_{ex} = 0$) and the case incorporating finite excess conductivities extracted from Table~\ref{tb:phil} for the silica and borosilicate substrates.
At moderate to high electrolyte concentrations ($C \gtrsim 1~\mathrm{mM}$), all configurations (Fig.~\ref{fig:dVdP}a-d) show close agreement between experimental data and model predictions, whether or not excess conductivity is considered. This consistency indicates that, under these conditions, electrical conductivity is governed predominantly by bulk and EDL ion transport, consistent with a flat-interface model.
At lower concentrations ($C \lesssim 1~\mathrm{mM}$), however, the measured electrokinetic coupling magnitude (Fig.~\ref{fig:dVdP}a-d) is significantly lower than predicted by models neglecting $\kappa_{ex}$. This discrepancy is most evident in the plain silica channel (Channel A), where $|dV/dp|$ saturates at approximately 6~mV/kPa for $C \lesssim 0.1$~mM and declines by orders of magnitude below the theoretical expectation as concentration decreases.
In contrast, the borosilicate glass channel (Channel D) exhibits smaller deviations from predictions without excess conductivity across the studied concentration range (Fig.~\ref{fig:dVdP}d), suggesting a weaker interfacial conduction effect.

To interpret results for the topographically modified Si channels (Channels B and C, Fig.~\ref{fig:dVdP}b-c), we employ Eq.~\ref{eq:kcontrol}, which includes the finite conductivity of liquid-filled cavities defined by Eq.~\ref{eq:kC}.
As illustrated in Fig.~\ref{fig:dVdP}, surface modifications in Channels B and C lead to enhanced pressure-to-voltage conversion, corresponding to reductions in excess interfacial conductivity as predicted by Eqs.~\ref{eq:kcontrol}-\ref{eq:kC}. Notably, the addition of a single 1-mm-wide cavity in Channel C yields a nearly tenfold increase in conversion relative to the unmodified Si surface in Channel A.
Figure~\ref{fig:dVdP}e summarizes the maximum pressure-to-voltage conversion achieved for each configuration, while Fig.~\ref{fig:dVdP}f shows the predicted excess conductivity in the limit of vanishing electrolyte concentration ($C \to 0$).
For the nanostructured Si surface in Channel B (50-nm-wide cavities), a 35\% increase in conversion is observed relative to Channel A, with peak performance at $C = 0.1$~mM, where the Debye length is approximately $\lambda_D \simeq 30$~nm.
Meanwhile, the predicted zero-concentration excess conductivity for Channel B is roughly 70\% greater than that of the plain Si surface (Fig.~\ref{fig:dVdP}f), a difference attributed to the accumulation of mobile protons in nanoscale cavities with high surface-to-volume ratios and sub-Debye dimensions at low concentrations ($C < 0.1$~mM).

\section{Conclusion}
\label{sec:conclusions}
Experimental measurements and analytical results reveal that a small but finite excess conductivity can substantially limit the efficiency of mechanical-to-electrical energy conversion in electrokinetic flows through microscale channels, particularly at low electrolyte concentrations.
When the concentration drops below 1~mM, this excess conductivity becomes the primary contributor to the total electrical conductivity and thus a critical constraint on energy conversion performance.
Under these low-concentration conditions, the Debye length exceeds 10~nm and becomes larger than the characteristic dimensions of the surface topography. As a result, nanoscale surface features become saturated with highly mobile protons released from the negatively charged substrate.
Notably, the excess conductivities measured in the present slit microchannels ($\kappa_{ex} \simeq 10^{-5}$--$10^{-3}$~S/m) are comparable in magnitude to those previously reported for slit nanochannels and nanopores at similarly low electrolyte concentrations. In such nanoconfined systems, these elevated conductivities have been attributed to charge-enhanced conduction phenomena and the high mobility of confined protons~\cite{stein2004surface,schoch2005ion,ritt2022thermodynamics,lee2012large,lin2021surface,zhan2023ion,aluru2023fluids}.

The analysis presented in this work demonstrates that the geometry and characteristic dimensions of micro- and nanoscale surface topography, together with the solid-phase electrical conductivity of the substrate, govern the magnitude of excess interfacial conductivity and thereby set the upper bound for pressure-to-voltage conversion in steady electrokinetic flow.
In agreement with previous studies, channels with native silica surfaces exhibit the highest excess conductivities. Various ion transport mechanisms have been proposed to explain this behavior, including conduction within the Stern layer and heterogeneities in surface charge distribution~\cite{lyklema1998surface,zhang2011water,werkhoven2018flow}.
Remarkably, our findings show that borosilicate glass channels exhibit significantly lower excess conductivities, on the order of $\kappa_{ex} \sim 10^{-5}$~S/m, nearly two orders of magnitude below those measured for silica despite having comparable nanoscale surface topography and surface chemistry.
This substantially reduced excess conductivity in borosilicate glass is attributed to its very low intrinsic solid-phase conductivity and its limited capacity to maintain a uniform surface potential $\psi_d$, in contrast to the more conductive silicon-based substrates.

The topographic modifications applied to silica surfaces known to exhibit substantial excess interfacial conductivities are shown to enhance pressure-to-voltage conversion, consistent with predictions from the proposed analytical model for excess conductivity.
To reduce this excess conductivity, we studied incorporating liquid-filled cavities that function as resistive elements in series, thereby restricting the transport of protons and ions along the interfacial conduction pathway. This strategy is well-suited for hydrophilic surface materials.
By enabling control over the EDL and the modulation of ion conduction, the topographic design approach presented here, combined with the developed analytical framework, provides practical guidance for engineering next-generation micro- and nanoscale fluidic systems aimed at improving mechanical-to-electrical energy conversion efficiency.
In particular, the integration of resistive, liquid-filled features such as nanoscale or micrometer-scale cavities with engineered surface-to-volume ratios introduces a tunable design parameter that can be optimized for specific electrolyte concentration regimes.

These results are highly relevant to a wide range of electrokinetic technologies, including self-powered microfluidic sensors, lab-on-chip platforms, ionic computing systems, and energy harvesting devices, all of which benefit from precise control over EDL structure, streaming potential, electroosmotic flow, and charge separation.
The experimental setup developed for this study, featuring custom-designed microchannels with diverse geometries and nanostructured surface modifications, also serves as a flexible testbed for future exploration of alternative substrate materials (e.g., doped semiconductors, polymer composites) and surface architectures.
The findings and methods reported in this work offer a scalable and adaptable approach for improving the performance and extending the applicability of electrokinetic devices across a broad spectrum of scientific and engineering domains.

\section{Acknowledgments}
The experimental analysis and theoretical model development in this work was supported by the Center for Mesoscale Transport Properties, an Energy Frontier Research Center funded by the U.S. Department of Energy, Office of Science, Basic Energy Sciences, under award DE-SC0012673.
A.\,D.\, and K.\,I.\ were supported by the National Science Foundation under award CBET-2417797.
This work employed resources of the Center for Functional Nanomaterials, which is a U.S. DOE Office of Science Facility, at Brookhaven National Laboratory under Contract DE-SC0012704.
%

\begin{thebibliography}{10}
\expandafter\ifx\csname url\endcsname\relax
  \def\url#1{\texttt{#1}}\fi
\expandafter\ifx\csname urlprefix\endcsname\relax\def\urlprefix{URL }\fi
\expandafter\ifx\csname href\endcsname\relax
  \def\href#1#2{#2} \def\path#1{#1}\fi

\bibitem{oldham1963streaming}
I.~Oldham, F.~Young, J.~Osterle, Streaming potential in small capillaries, J.
  Colloid Sci. 18~(4) (1963) 328--336.

\bibitem{osterle1964}
J.~Osterle, Electrokinetic energy conversion, J. Appl. Mech. 31~(2) (1964)
  161--164.

\bibitem{morrison1965electrokinetic}
F.~Morrison~Jr, J.~Osterle, Electrokinetic energy conversion in ultrafine
  capillaries, J. Chem. Phys. 43~(6) (1965) 2111--2115.

\bibitem{schoch2008}
R.~B. Schoch, J.~Han, P.~Renaud, Transport phenomena in nanofluidics, Rev. Mod.
  Phys. 80~(3) (2008) 839--883.

\bibitem{eijkel2005}
J.~C. Eijkel, A.~v.~d. Berg, Nanofluidics: what is it and what can we expect
  from it?, Microfluid. Nanofluidics 1 (2005) 249--267.

\bibitem{pennathur2005}
S.~Pennathur, J.~G. Santiago, Electrokinetic transport in nanochannels. 1.
  theory, Anal. Chem. 77~(21) (2005) 6772--6781.

\bibitem{haldrup2016tailor}
S.~Haldrup, J.~Catalano, M.~Hinge, G.~V. Jensen, J.~S. Pedersen, A.~Bentien,
  Tailoring membrane nanostructure and charge density for high electrokinetic
  energy conversion efficiency, ACS Nano 10~(2) (2016) 2415--2423.

\bibitem{qin2020constant}
Y.~Qin, Y.~Wang, X.~Sun, Y.~Li, H.~Xu, Y.~Tan, Y.~Li, T.~Song, B.~Sun, Constant
  electricity generation in nanostructured silicon by evaporation-driven water
  flow, Angew. Chem. 132~(26) (2020) 10706--10712.

\bibitem{aktar2020}
A.~Al~Hossain, M.~Yang, A.~Checco, G.~Doerk, C.~E. Colosqui, Large-area
  nanostructured surfaces with tunable zeta potentials, Appl. Mater. Today 19
  (2020) 100553.

\bibitem{wang2022mxenes}
Y.~Wang, T.~Guo, Z.~Tian, K.~Bibi, Y.-Z. Zhang, H.~N. Alshareef, Mxenes for
  energy harvesting, Adv. Mater. 34~(21) (2022) 2108560.

\bibitem{zhang2017capillary}
X.~Zhang, H.~Liu, C.~Liu, K.~Li, X.~Li, Capillary driven electrokinetic
  generator for environmental energy harvesting, Mater. Res. Bull. 89 (2017)
  9--14.
\newblock \href {https://doi.org/10.1016/j.materresbull.2017.01.013}
  {\path{doi:10.1016/j.materresbull.2017.01.013}}.

\bibitem{jiao2023}
K.~Jiao, H.~Yan, F.~Qian, W.~Zhang, H.~Li, Q.~Wang, C.~Zhao, Energy harvesting
  based on water evaporation-induced electrokinetic streaming potential/current
  in porous carbonized carrots, J. Power Sources 569 (2023) 233007.

\bibitem{liu2024energy}
Z.~Liu, Q.~Wang, T.~Chen, K.~Wang, G.~Liu, Energy harvesting from carbon-based
  rope driven by capillary flow, J. Power Sources 618 (2024) 235193.

\bibitem{yuan2024recent}
Z.~Yuan, L.~Guo, Recent advances in solid--liquid triboelectric nanogenerator
  technologies, affecting factors, and applications, Sci. Rep. 14~(1) (2024)
  10456.

\bibitem{zhao2024design}
K.~Zhao, Z.~Gao, J.~Zhang, J.~Zhou, F.~Zhan, L.~Qiang, M.-J. Liu, R.-H. Cyu,
  Y.-L. Chueh, Design of strong-performance, high-heat dissipation rate, and
  long-lifetime triboelectric nanogenerator based on robust hexagonal boron
  nitride (hbn) nanosheets/polyvinyl chloride (pvc) composite films for
  rotational energy harvesting, J. Power Sources 614 (2024) 234997.

\bibitem{somton2025transparent}
S.~Somton, S.~Wanwong, W.~Sangkhun, N.~Poolthong, Transparent
  perfluorodecyltriethoxysilane (pfdtes)/pdms based water droplet-driven
  triboelectric nanogenerators as self-powered energy devices, J. Power Sources
  625 (2025) 235693.

\bibitem{iglesias2020combining}
G.~Iglesias, S.~Ahualli, A.~Delgado, P.~Arenas-Fernandez, M.~Fernandez,
  Combining soft electrode and ion exchange membranes for increasing salinity
  difference energy efficiency, J. Power Sources 453 (2020) 227840.

\bibitem{wang2024salinity}
Y.~Wang, L.~Zhao, M.~Chen, Q.~Liu, T.~Zhang, Salinity gradient induced blue
  energy generation using two-dimensional materials, Nat. Commun. 15~(1) (2024)
  4567.
\newblock \href {https://doi.org/10.1038/s41467-024-04567-8}
  {\path{doi:10.1038/s41467-024-04567-8}}.

\bibitem{ma2019energy}
J.~Ma, P.~Liang, X.~Sun, H.~Zhang, Y.~Bian, F.~Yang, J.~Bai, Q.~Gong, X.~Huang,
  Energy recovery from the flow-electrode capacitive deionization, J. Power
  Sources 421 (2019) 50--55.

\bibitem{younes2024review}
H.~Younes, D.~Lou, M.~Mao, M.~M. Rahman, M.~AlNahyan, H.~Younis, H.~Hong, M.~K.
  Datta, A review on capacitive deionization: Recent advances in prussian blue
  analogues and carbon materials based electrodes, Hybrid Advances (2024)
  100191.

\bibitem{grover2013preparation}
I.~S. Grover, S.~Singh, B.~Pal, The preparation, surface structure, zeta
  potential, surface charge density and photocatalytic activity of tio2
  nanostructures of different shapes, Appl. Surf. Sci. 280 (2013) 366--372.

\bibitem{huang2020molecular}
S.~Huang, A.~M. Rahmani, T.~Singletary, C.~E. Colosqui, Molecular dynamics and
  continuum analyses of the electrokinetic zeta potential in nanostructured
  slit channels, Colloids Surf. A 603 (2020) 125100.

\bibitem{goyal2024generalizing}
V.~Goyal, S.~Datta, S.~Chakraborty, Generalizing electroosmotic-flow
  predictions over charge-modulated periodic topographies: tuneable far-field
  effects, J. Fluid Mech. 990 (2024) A1.

\bibitem{onsager1931reciprocal}
L.~Onsager, Reciprocal relations in irreversible processes. i., Phys. Rev.
  37~(4) (1931) 405.

\bibitem{miller1960thermo}
D.~G. Miller, Thermodynamics of irreversible processes. the experimental
  verification of the onsager reciprocal relations., Chem. Rev. 60~(1) (1960)
  15--37.

\bibitem{xuan2006thermodynamic}
X.~Xuan, D.~Li, Thermodynamic analysis of electrokinetic energy conversion, J.
  Power Sources 156~(2) (2006) 677--684.

\bibitem{van2006electrokinetic}
F.~H. Van~der Heyden, D.~J. Bonthuis, D.~Stein, C.~Meyer, C.~Dekker,
  Electrokinetic energy conversion efficiency in nanofluidic channels, Nano
  Lett. 6~(10) (2006) 2232--2237.

\bibitem{joly2004hydrodynamics}
L.~Joly, C.~Ybert, E.~Trizac, L.~Bocquet, Hydrodynamics within the electric
  double layer on slipping surfaces, Phys. Rev. Lett. 93~(25) (2004) 257805.

\bibitem{davidson2008electrokinetic}
C.~Davidson, X.~Xuan, Electrokinetic energy conversion in slip nanochannels, J.
  Power Sources 179~(1) (2008) 297--300.

\bibitem{ren2008slip}
Y.~Ren, D.~Stein, Slip-enhanced electrokinetic energy conversion in nanofluidic
  channels, Nanotechnol. 19~(19) (2008) 195707.

\bibitem{papadopoulos2012electrokinetics}
P.~Papadopoulos, X.~Deng, D.~Vollmer, H.-J. Butt, Electrokinetics on
  superhydrophobic surfaces, J. Condens. Matter Phys. 24~(46) (2012) 464110.

\bibitem{berli2010electrokinetic}
C.~L. Berli, Electrokinetic energy conversion in microchannels using polymer
  solutions, J. Colloid Interface Sci. 349~(1) (2010) 446--448.

\bibitem{zhao2013electrokinetics}
C.~Zhao, C.~Yang, Electrokinetics of non-newtonian fluids: a review, Adv.
  Colloid Interface Sci. 201 (2013) 94--108.

\bibitem{zhang2020soft}
J.~Zhang, K.~Zhan, S.~Wang, X.~Hou, Soft interface design for electrokinetic
  energy conversion, Soft Matter 16~(12) (2020) 2915--2927.

\bibitem{joly2006liquid}
L.~Joly, C.~Ybert, E.~Trizac, L.~Bocquet, Liquid friction on charged surfaces:
  From hydrodynamic slippage to electrokinetics, J. Chem. Phys. 125~(20)
  (2006).

\bibitem{heverhagen2016slip}
J.~Heverhagen, M.~Tasinkevych, A.~Rahman, C.~T. Black, A.~Checco, Slip length
  enhancement in nanofluidic flow using nanotextured superhydrophobic surfaces,
  Adv. Mater. Interfaces 3~(17) (2016) 1600303.

\bibitem{stein2004surface}
D.~Stein, M.~Kruithof, C.~Dekker, Surface-charge-governed ion transport in
  nanofluidic channels, Phys. Rev. Lett. 93~(3) (2004) 035901.

\bibitem{schoch2005ion}
R.~B. Schoch, P.~Renaud, Ion transport through nanoslits dominated by the
  effective surface charge, Appl. Phys. Lett. 86~(25) (2005).

\bibitem{ritt2022thermodynamics}
C.~L. Ritt, J.~P. de~Souza, M.~G. Barsukov, S.~Yosinski, M.~Z. Bazant, M.~A.
  Reed, M.~Elimelech, Thermodynamics of charge regulation during ion transport
  through silica nanochannels, ACS Nano 16~(9) (2022) 15249--15260.

\bibitem{lee2012large}
C.~Lee, L.~Joly, A.~Siria, A.-L. Biance, R.~Fulcrand, L.~Bocquet, Large
  apparent electric size of solid-state nanopores due to spatially extended
  surface conduction, Nano Lett. 12~(8) (2012) 4037--4044.

\bibitem{lin2021surface}
K.~Lin, Z.~Li, Y.~Tao, K.~Li, H.~Yang, J.~Ma, T.~Li, J.~Sha, Y.~Chen, Surface
  charge density inside a silicon nitride nanopore, Langmuir 37~(35) (2021)
  10521--10528.

\bibitem{zhan2023ion}
L.~Zhan, Z.~Zhang, F.~Zheng, W.~Liu, Y.~Zhang, J.~Sha, Y.~Chen, Ion
  concentration-dependent surface charge density inside a nanopore, J. Phys.
  Chem. Letters 14~(50) (2023) 11536--11542.

\bibitem{aluru2023fluids}
N.~R. Aluru, F.~Aydin, M.~Z. Bazant, D.~Blankschtein, A.~H. Brozena, J.~P.
  de~Souza, M.~Elimelech, S.~Faucher, J.~T. Fourkas, V.~B. Koman, et~al.,
  Fluids and electrolytes under confinement in single-digit nanopores, Chem.
  Rev. 123~(6) (2023) 2737--2831.

\bibitem{healy1978}
T.~W. Healy, L.~R. White, Ionizable surface group models of aqueous interfaces,
  Adv. Colloid Interface Sci. 9~(4) (1978) 303--345.

\bibitem{grahame1947electrical}
D.~C. Grahame, The electrical double layer and the theory of
  electrocapillarity., Chem. Rev. 41~(3) (1947) 441--501.

\bibitem{duan2010anomalous}
C.~Duan, A.~Majumdar, Anomalous ion transport in 2-nm hydrophilic nanochannels,
  Nat. Nanotechnol. 5~(12) (2010) 848--852.

\bibitem{paiva2019conduction}
V.~T. Paiva, L.~P. Santos, D.~S. da~Silva, T.~A. Burgo, F.~Galembeck,
  Conduction and excess charge in silicate glass/air interfaces, Langmuir
  35~(24) (2019) 7703--7712.

\bibitem{ji2021electric}
A.~Ji, Y.~Chen, Electric control of ionic transport in sub-nm nanopores, RSC
  Adv. 11~(23) (2021) 13806--13813.

\bibitem{allemand2023anomalous}
A.~Allemand, M.~Zhao, O.~Vincent, R.~Fulcrand, L.~Joly, C.~Ybert, A.-L. Biance,
  Anomalous ionic transport in tunable angstrom-size water films on silica,
  Proc. Natl. Acad. Sci. 120~(25) (2023) e2221304120.

\bibitem{dhiraj2020}
D.~Nandyala, Z.~Wang, D.~Hwang, T.~Cubaud, C.~E. Colosqui, Design, fabrication,
  and analysis of a capillary diode for potential application in water--oil
  separation, ACS Appl. Mater. Interfaces 12~(41) (2020) 45950--45960.

\bibitem{zhen2021}
Z.~Wang, D.~Nandyala, C.~E. Colosqui, T.~Cubaud, D.~J. Hwang, Glass surface
  micromachining with simultaneous nanomaterial deposition by picosecond laser
  for wettability control, Appl. Surf. Sci. 546 (2021) 149050.

\bibitem{aktar2022}
A.~Al~Hossain, A.~Dick, G.~Doerk, C.~E. Colosqui, Toward controlling wetting
  hysteresis with nanostructured surfaces derived from block copolymer
  self-assembly, Nanotechnology 33~(45) (2022) 455302.

\bibitem{agmon1995grotthuss}
N.~Agmon, The grotthuss mechanism, Chem. Phys. Lett. 244~(5-6) (1995) 456--462.

\bibitem{kornyshev2003}
A.~Kornyshev, A.~Kuznetsov, E.~Spohr, J.~Ulstrup, Kinetics of proton transport
  in water, J. Phys. Chem. B 107~(15) (2003) 3351--3366.

\bibitem{popov2023}
I.~Popov, Z.~Zhu, A.~R. Young-Gonzales, R.~L. Sacci, E.~Mamontov, C.~Gainaru,
  S.~J. Paddison, A.~P. Sokolov, Search for a grotthuss mechanism through the
  observation of proton transfer, Commun. Chem. 6~(1) (2023) 77.

\bibitem{behrens2001charge}
S.~H. Behrens, D.~G. Grier, The charge of glass and silica surfaces, J. Chem.
  Phys. 115~(14) (2001) 6716--6721.

\bibitem{barisik2014size}
M.~Barisik, S.~Atalay, A.~Beskok, S.~Qian, Size dependent surface charge
  properties of silica nanoparticles, J. Phys. Chem. C 118~(4) (2014)
  1836--1842.

\bibitem{yang2020understanding}
J.~Yang, H.~Su, C.~Lian, Y.~Shang, H.~Liu, J.~Wu, Understanding surface charge
  regulation in silica nanopores, Phys. Chem. Chem. Phys. 22~(27) (2020)
  15373--15380.

\bibitem{lyklema1998surface}
J.~Lyklema, M.~Minor, On surface conduction and its role in electrokinetics,
  Colloids Surf. A 140~(1-3) (1998) 33--41.

\bibitem{zhang2011water}
H.~Zhang, A.~A. Hassanali, Y.~K. Shin, C.~Knight, S.~J. Singer, The
  water--amorphous silica interface: analysis of the stern layer and surface
  conduction, J. Chem. Phys. 134~(2) (2011) 024705.

\bibitem{werkhoven2018flow}
B.~Werkhoven, J.~C. Everts, S.~Samin, R.~Van~Roij, Flow-induced surface charge
  heterogeneity in electrokinetics due to stern-layer conductance coupled to
  reaction kinetics, Phys. Rev. Lett. 120~(26) (2018) 264502.

\end{thebibliography}

\end{document}